\documentclass[]{aastex631}
\usepackage{amsmath}
\usepackage{float}
\usepackage{graphicx}
\usepackage{subfigure}
\usepackage{txfonts}
\usepackage{indentfirst}
\usepackage{ulem}
\usepackage[Symbolsmallscale]{upgreek}
\usepackage{nomencl}
\usepackage{enumitem}

\newcommand{\eq}[1]{\begin{equation}  #1 \end{equation}}

\newcommand{\bb}[1]{\left[ #1 \right]}

\newcommand{\dd}{{\rm d}}
\newcommand{\expo}[1]{~{\rm e}^{ #1 }}

\newcommand{\ic}{{\rm i}}

\def\apj{ApJ}

\def\mnras{MNRAS}
\def\nat{Nature}

\def\prd{Phys.~Rev.~D}

\begin{document}
\title{Stokes phenomena in lensing}
 
\author[0000-0003-2076-4510]{Xun Shi}
\affiliation{South-Western Institute for Astronomy Research (SWIFAR), \\
Yunnan University, 650500 \\
Kunming, P. R. China}

\begin{abstract} 
As lensing of coherent astrophysical sources e.g. pulsars, fast radio bursts, and gravitational waves becomes observationally relevant, 
the mathematical framework of Picard-Lefschetz theory has recently been introduced to fully account for wave optics effects. 
Accordingly, the concept of lensing images has been generalized to include complex solutions of the lens equation referred to as ``imaginary images'', and more radically, to the Lefschetz thimbles which are a sum of steepest descent contours connecting the real and imaginary images in the complex domain.
In this wave-optics-based theoretical framework of lensing, we study the ``Stokes phenomena'' as the change of the topology of the Lefschetz thimbles. 
Similar to the well-known caustics at which the number of geometric images changes abruptly, the corresponding Stokes lines are the boundaries in the parameter space where the number of effective imaginary images changes.
We map the Stokes lines for a few lens models. 
The resulting Stokes line-caustics network represents a unique feature of the lens models. 
The observable signature of the Stokes phenomena is the change of interference behavior, in particular the onset of frequency oscillation for some Stokes lines.  
We also demonstrate high-order Stokes phenomena where the system has a continuous number of effective images but with an abrupt change in the way they are connected to each other by the Lefschetz thimbles.
Their full characterization calls for an analogy of the catastrophe theory for caustics.
\end{abstract}



\section{Introduction}

Traditional study of astrophysical lensing has been based on the ray optics picture, which is valid for most astrophysical lensing situations \citep{schneider92}.
In this regime, the lensing effect can be described by a set of geometric images, which are the solutions of the lens equation with real positions.
The wave effect can be included to the first order by considering the interference of these geometric images, resulting in the eikonal approximation of lensing which is widely used in the field of plasma lensing and/or pulsar scintillation.

Nevertheless, for a variety of lens models, solutions of the lensing equation with non-zero imaginary parts can be found. 
They are referred to as the ``imaginary images''. 
\citet{grillo18} and \citet{jow21} have shown that near caustics where geometric and eikonal approximations fail, imaginary images can contribute significantly to the distribution of light.
\citet{jow21} further argue that, in the weak lensing regime where only one geometric image is present, the imaginary images can introduce interference which is
 potentially observable for coherent sources and useful for inferring lens parameters.

In accordance with the imaginary images, the concept of ``Stokes phenomena'' has been introduced \citep{jow21}.
Generically, the locations of the lensing images vary continuously with the control parameters of the lens equation.
However, at certain boundaries in the parameter space, an annihilation/creation of images occurs.
Whereas the boundaries at which geometric images annihilate into imaginary images are referred to as caustics, the boundaries at which the number of effective imaginary images changes are referred to as the Stokes lines, and the associated Stokes line crossing referred to as the Stokes phenomena.
Stokes lines, like the caustics, are characteristic features associated with lens models. 
Thus, it is of theoretical interest to map the Stokes lines for a variety of lens models.  

Although the concept of imaginary images and Stokes phenomena exist in the eikonal limit, they are actually based on a full description of the wave optics lensing with the Picard-Lefschetz theory (see \citealt{witten10} for a description)
which has recently been introduced to evaluate diffraction integrals in the wave optics formulation lensing by \citet{feldbrugge19,feldbrugge23}.
The Picard-Lefschetz theory generalizes the concept of geometric and imaginary images into the pieces of the Lefschetz thimbles associated with them \citep{jow21}.
It is crucial in both conceptually understanding the imaginary images and the Stokes phenomena, and in practically determining them.
Most importantly, the Picard-Lefschetz theory offers a way to compute the electric field at low frequencies where the wave effects are significant, 
which has direct application in lensing of coherent sources e.g. 
pulsars, fast radio bursts, and gravitational waves \citep[see][and references therein]{cordes17, main18, jow20, lin21, suvorov22, caliskan23, leung23, savastano23, tambalo23, zhu23}.

In this paper, we systematically study the Stokes phenomena using the Picard-Lefschetz theory which provides a more fundamental definition of the Stokes phenomena as the change of the topology of the Lefschetz thimbles.
We shall provide examples of Stokes line networks for a few simple lens models, and how the number of effective imaginary images changes in the parameter space.
We study how the electric field changes across the Stokes line at various observing frequencies, and identify possible observable signatures of the Stokes phenomena.
Apart from first-order Stokes phenomena where the number of effective imaginary images is not continuous, we propose the existence of high-order Stokes phenomena where the system has continuous imaginary images but with an abrupt change in the way they are connected to each other by the Lefschetz thimbles.
As an initial effort, we limit ourselves to 1D situations.


\begin{figure}
    \centering
    \includegraphics[width=0.38\textwidth]{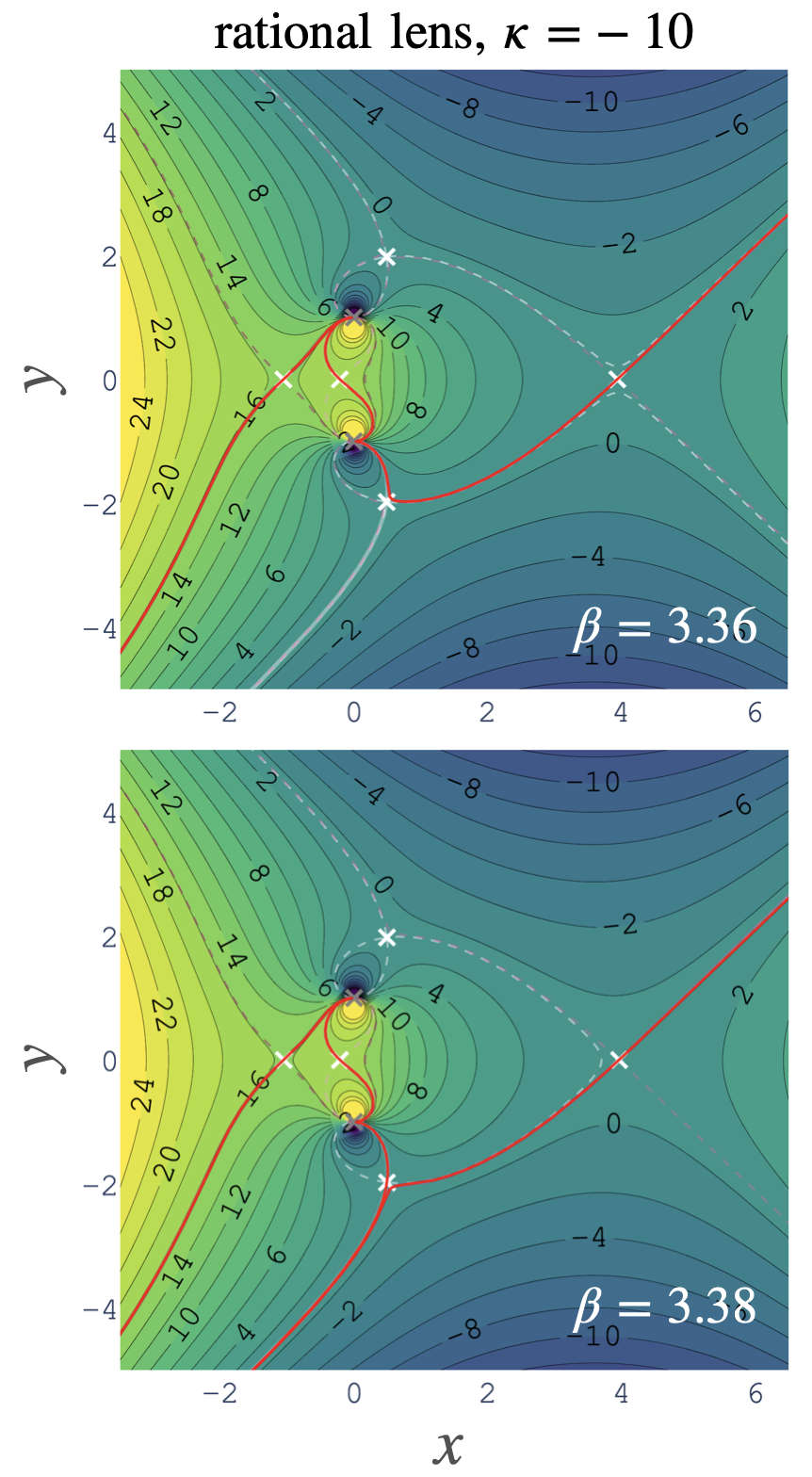}
        \caption{
            \textbf{Example of the Stokes phenomena} as an abrupt change of the topology of the Lefschetz thimbles (red solid lines).
            This particular example shows the Stokes phenomena occurring at a lens amplitude $\kappa=-10$ for the converging rational lens between source locations $\beta = 3.36$ and $\beta = 3.38$. 
            The Lefschetz thimbles are a subset of the steepest descending and ascending contours (dashed lines) of the $h$ function that are connected to the stationary phase points (white crosses).
            Each piece of Lefschetz thimble is associated with a geometric or imaginary image (white crosses on the red lines), and end at either an infinity or a pole (gray crosses).
            Along a piece of Lefschetz thimble, the $H$ function (background contours) is constant.   
        } 
        \label{fig:stokes_demo} 
\end{figure}

\begin{figure}
    \centering
    \includegraphics[width=0.4\textwidth]{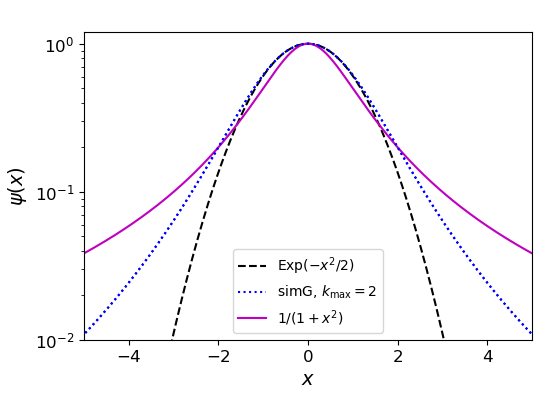}     
        \caption{
          Lens shapes used in this paper: a rational lens (magenta line) and a $k_{\rm max}=2$ simG lens (blue dotted line).
          The Gaussian lens shape is shown as the black dashed line as a comparison.
        }
        \label{fig:psi} 
\end{figure} 

\begin{figure}
    \centering
    \includegraphics[width=0.4\textwidth]{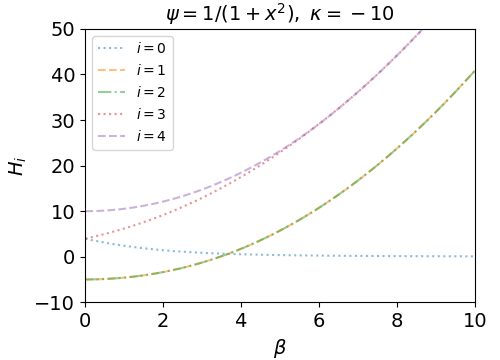
    }
        \caption{
            Coincidence of the $H$ values of different images is used to identify the locations of Stokes lines.
            In this example of the converging rational lens with an amplitude $\kappa=-10$, there are 5 stationary phase points ($i=0$ to 4). 
            A Stokes line lies at the crossing of $H_0$ and $H_{1,2}$ at $\beta=3.67$, and a caustic lies at $\beta=5.93$ where $H_{3}$ and $H_{4}$ converge.
        } 
        \label{fig:find_stokes} 
\end{figure}

\begin{figure}
    \centering
    \includegraphics[width=0.48\textwidth]{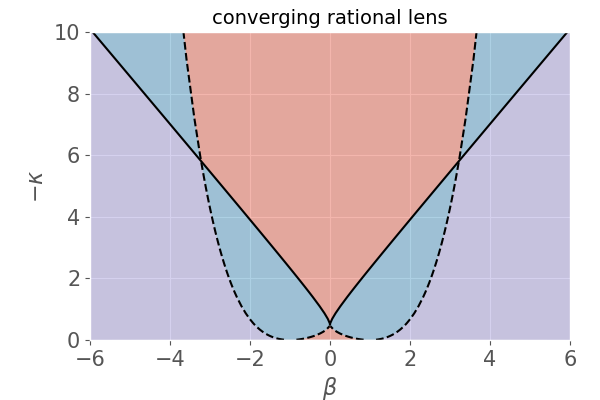}\\
    \includegraphics[width=0.48\textwidth]{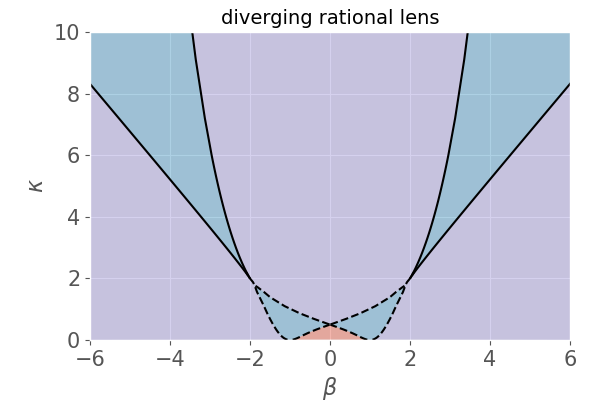}\\
    \includegraphics[width=0.48\textwidth]{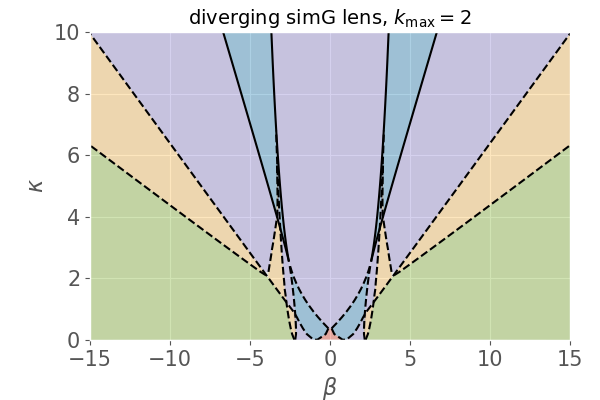}
        \caption{
            \textbf{Stokes line} (black dashed lines) \textbf{- caustics} (black solid lines) \textbf{network}  on the parameter space spanned by lens amplitude $\kappa$ and its angular offset from the source location $\beta$. 
            The lens shapes are the converging rational lens (upper panel), diverging rational lens (middle panel), and the diverging simG lens (lower panel).
            The number of effective imaginary images in a region in this parameter space is indicated with the color of the background: red for zero, blue for one, purple for two, yellow for three, and green for four.
        } 
        \label{fig:stokes2} 
\end{figure}

\begin{figure*}
    \centering
    \includegraphics[width=0.9\textwidth]{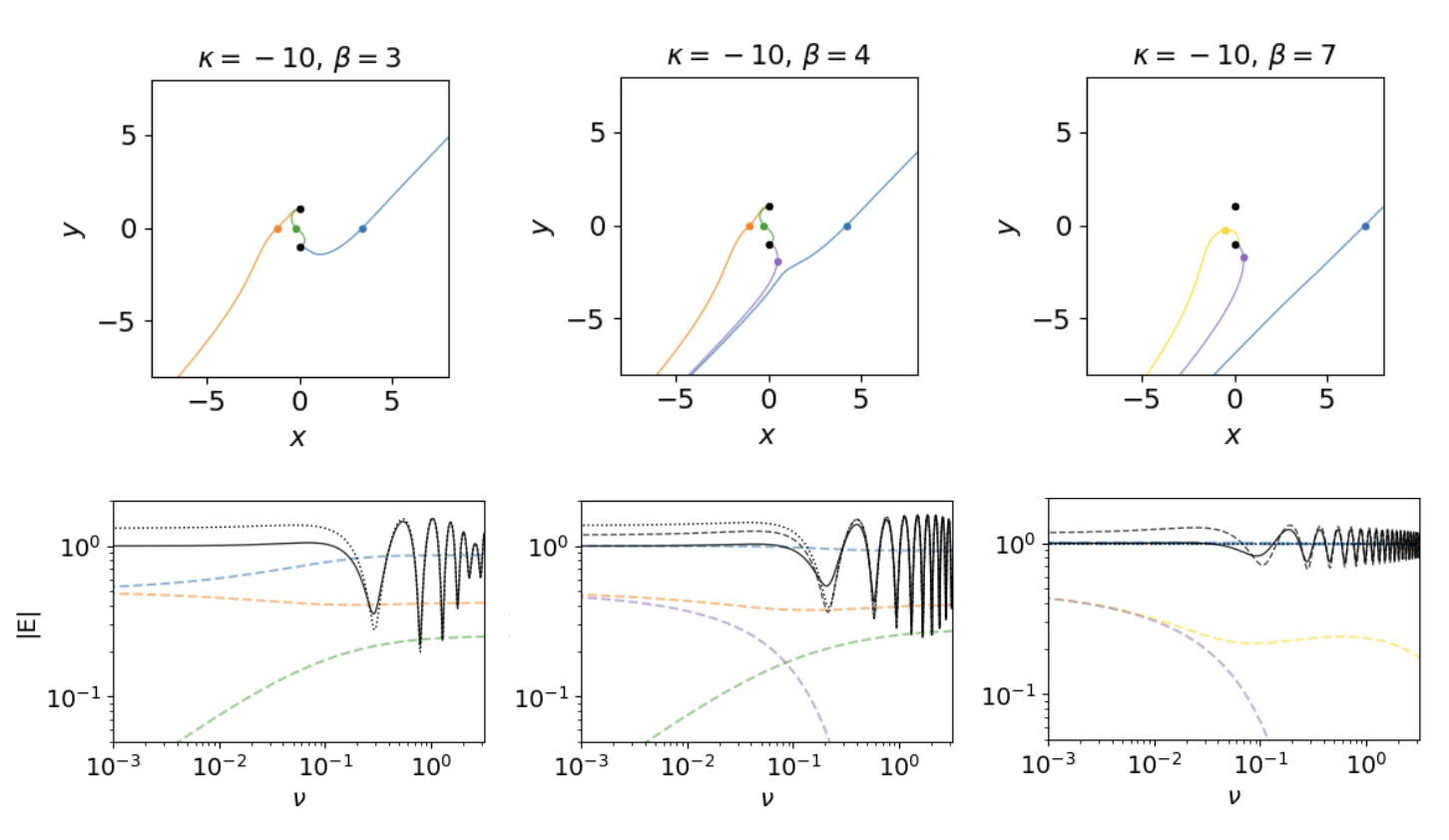}
    \caption{
        \textbf{Example of Stokes phenomena} (between the 1st and 2nd columns) \textbf{in comparison to caustic crossing} (between the 2nd and 3rd columns). 
        Shown are the Lefschetz thimbles (upper panels) and the amplitude of the electric field as a function of reduced frequency (lower panels) for a converging lens $\psi=1/(1+x^2)$ with amplitude $\kappa=-10$ and for three different source locations $\beta=3,4$ and 7.
        For this lensing system, there is a Stokes line at $x=3.67$ and a caustic at $x=5.93$ (see Fig.\;\ref{fig:stokes2}).
        The Stokes line crossing leads to an additional effective imaginary image (purple dot) for $\beta=4$ compared to $\beta=3$, and the caustic crossing leads to the annihilation of real images (orange and green dots) and the creation of one effective image (yellow dot) for $\beta=7$ compared to $\beta=4$.
        The electric fields are computed using the Picard-Lefschetz theory and are decomposed into contributions from Lefschetz thimble pieces associated with each image (color dashed lines in the bottom panels, with the colors matching with the upper panels).
        The total electric fields are shown as the solid black lines in the bottom panels, along with their eikonal approximation with imaginary image contribution (black dashed lines) and without imaginary image contribution (black dotted lines).
        }
    \label{fig:example}
\end{figure*}

\begin{figure}
    \centering
    \includegraphics[width=0.49\textwidth]{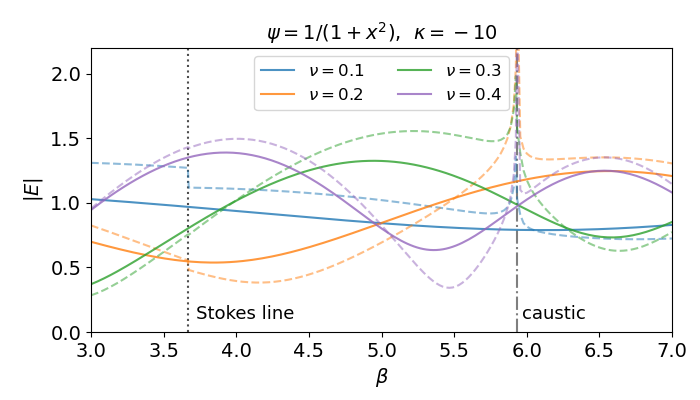}\\
    \includegraphics[width=0.49\textwidth]{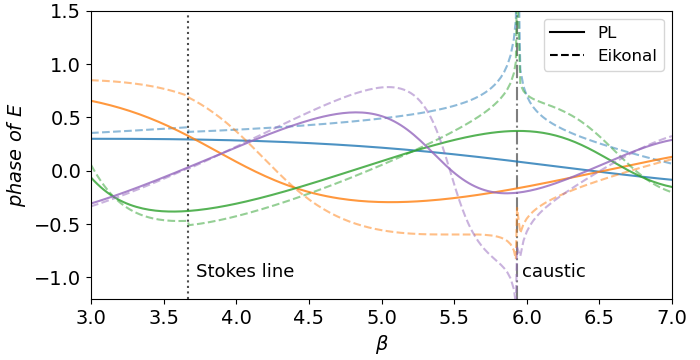}
    \caption{
        \textbf{Continuity of the electric field amplitude} (upper panel) \textbf{and phase} (lower panel) \textbf{across Stokes line and caustic.}  
        The lensing system is the same as Fig.\;\ref{fig:example}.
        Both the actual electric field computed using the Picard-Lefschetz theory (solid lines) and its eikonal approximation (dashed lines, with imaginary image contribution) are shown for four reduced frequency values.
    }
    \label{fig:continuity}
\end{figure}

\begin{figure}
    \centering
    \includegraphics[width=0.49\textwidth]{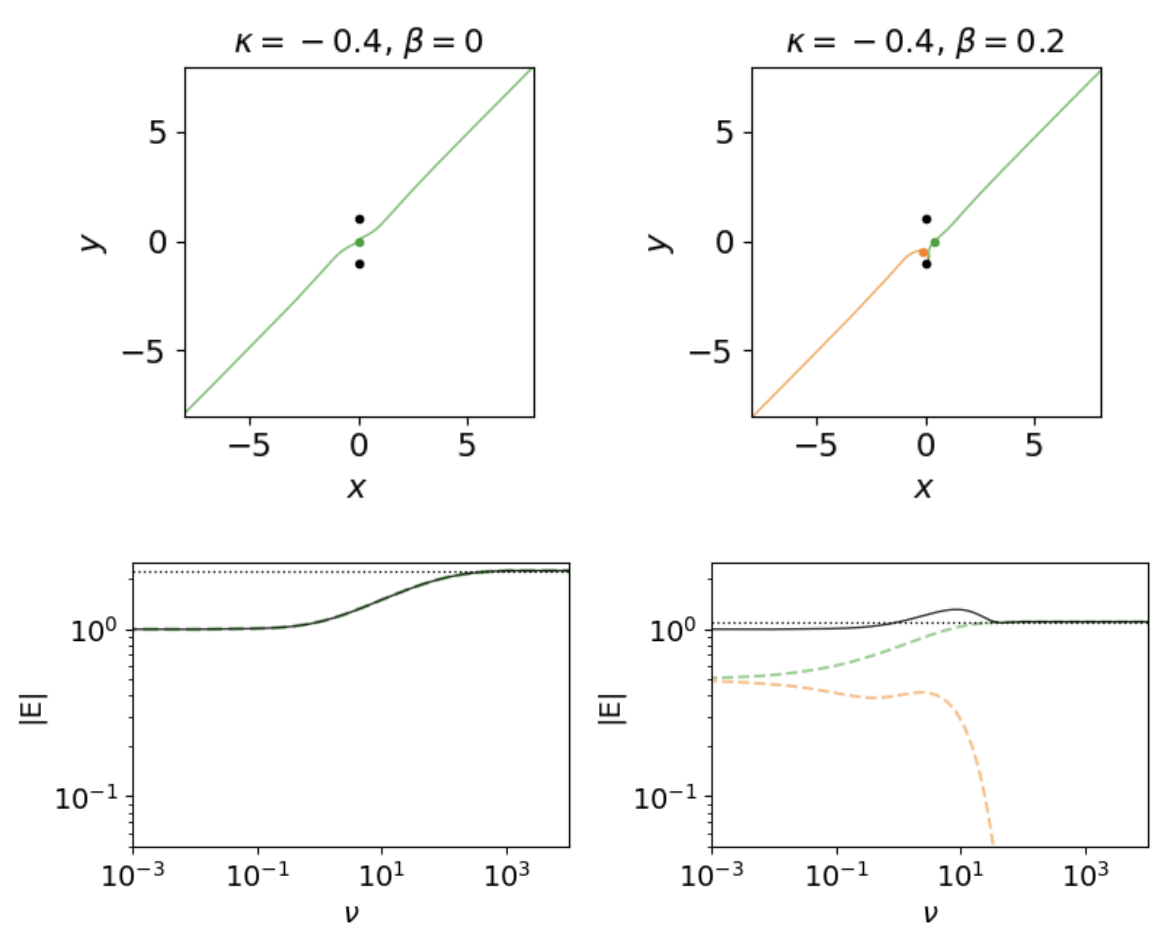}
    \caption{
      \textbf{Possible observable signature of Stokes phenomena: onset of frequency oscillation in the electric field.}
      This example shows the Lefschetz thimbles (upper panels) and the amplitude of the electric field as a function of reduced frequency (lower panels) for the converging rational lens with $\kappa=-0.4$ at $\beta=0$ (left panel) and $\beta=0.2$ (right panel).
      In the lower panel, the black solid and dotted line represent the total electric fields computed with the Picard-Lefschetz theory and their eikonal approximations, respectively. 
      The individual contributions from Lefschetz thimble pieces associated with each image are shown with the color dashed lines with colors matching the upper panels.
      The frequency oscillation in the bottom right panel is caused by the interference between the real image (green) and the imaginary image (orange) generated at a Stokes line.
    }
    \label{fig:sign_of_stokes}
\end{figure}

\section{Picard-Lefschetz theory and the imaginary images}
We start with a description of the wave optics formulation of lensing and the Picard-Lefschetz theory. 
They are crucial in both understanding the imaginary images and Stokes phenomena, and in determining which imaginary images are the effective ones as well as the locations of the Stokes lines.

\subsection{Wave optics formulation of lensing}
In the presence of a lens (or equivalently, a phase screen), the wave amplitude received by an observer for a point source with a unit magnitude is given by the Kirchhoff-Fresnel diffraction integral \citep[e.g.][]{schneider92, born}
\eq{
    E(\beta) = \sqrt{\frac{\nu}{2\uppi \ic}} \int_{-\infty}^{\infty} \exp\bb{\ic \nu \phi} \dd x \,.
}

This expression of the wave amplitude $E({\beta})$, also referred to as the `transmission factor' of a lensing configuration, describes the diffraction of light at an intermediate phase screen by summing over all possible wave paths connecting the source, the observer, and a point ${x}$ on the phase screen according to the Principle of Huygens.
Here, ${\beta}$ is the location of the source line-of-sight on the phase screen. 
The coordinates ${x}$ and ${\beta}$ are made dimensionless by scaling with the size of the lens $a_{\rm lens}$ following the convention of \citet{shi21} and \citet{jow23}.  
The reduced frequency $\nu$ is the key parameter that determines the importance of the wave effect. 
It is the square of the lens size $a_{\rm lens}$ - Fresnel scale $r_{\rm F}$ ratio which combines the observing wavelength $\lambda$ and the lensing distances \citet{jow23},
\eq{
    {\nu} = \frac{2\uppi{a_{\rm lens}}^2}{\lambda \bar{d}} = \frac{a_{\rm lens}^2}{r^2_{\rm F}} \,,
}
where $\bar{d} = D s (1-s)$, $D$ is the source distance, and $s$ is the fractional distance of the phase screen to the source. 
The Fresnel scale is defined as $r_{\rm F}^2 = \lambda \bar{d} / 2\uppi$.   


The phase function 
\eq{
    \phi(x; \beta, \kappa) = \frac{(x-\beta)^2}{2} - \kappa\psi(x)
}
accounts for the relative phases of the wave paths connecting the source, different locations $x$ on the lens plane, and the observer.
It is composed of both the geometrical phase delay and the dispersive phase delay introduced by the lens. 
Here, the lens is characterized with an amplitude $\kappa$ and a shape described by a positive function $\psi(x)$, and 
we have adopted the convention that a converging lens (e.g. a gravitational lens) has a negative $\kappa$ value. 

\subsection{Picard-Lefschetz theory}
The Picard-Lefschetz theory is an important tool in complex geometry and has been recently introduced to evaluate diffraction integrals in lensing by \citet{feldbrugge19,feldbrugge23}.  
By generalizing the integrand to the complex plane $x \to z = x + \ic y$, it provides an exact approach to computing highly oscillatory integrals for a wide range of lenses. 

The Picard-Lefschetz theory offers the benefit of maintaining the notion of a finite collection of images even at low frequencies when geometric optics fail. 
This is achieved by breaking the ultimate integration contour on the complex plane into distinct sub-contours known as `Lefschetz thimbles', each of which corresponds to a specific geometric image (see Fig.\;\ref{fig:stokes_demo} for an example). 
In addition to these Lefschetz thimbles associated with geometric images, there will generally be contributions from imaginary images, which are typically disregarded in the context of geometric optics.


A key procedure of the Picard-Lefschetz method is to decompose the phase function into real and imaginary parts,
\eq{
    \ic \phi \equiv h + \ic H
}
where $h$ and $H$ are real functions on the complex plane. 
By definition, the Lefschetz thimbles are the steepest descent contours of the $h$-function.
Therefore, along a piece of Lefschetz thimble, the $H$-function is constant, and the $h$-function decreases rapidly from the image locations.

\subsection{Imaginary images}
Traditionally, it has been taken for granted that only the real solutions of the lens equation, referred to as the `geometric images', would contribute to the signal received by the observer. 
Nonetheless, the lens equation can yield complex solutions that have non-zero imaginary parts. 
Some of these so-called imaginary images can contribute to the electric field at the observer, and are referred to as the `effective imaginary images'.
The effective imaginary images can be identified by the fact that their steepest ascent contours of the $h$-function have none-zero intersection with the real axis.
They can have a significant impact on the total electric field in specific circumstances \citep{jow21},
and should be taken into account in the eikonal limit as well. 

In the eikonal limit, the field associated with an image, either real or imaginary, is \citep{conner73, grillo18}
\eq{
    E_i = \frac{\expo{\ic \nu \phi_i}}{\Delta_i^{1/2}}
    \label{eq:Ei}
}
with subscription $i$ indicating evaluation at image location ${z}_i$, and
\eq{
    \Delta = \rm{det} \frac{\partial^2 \phi}{\partial {x}^2} \,.
}
Note that $\Delta$ is complex, and the square root function in the denominator of Eq.\;\ref{eq:Ei} is the usual branch of the analytically continued square root function. Thus, the Morse phases have been included in $\Delta^{1/2}$.

The magnifications of geometric images are \citep{schneider92}
\eq{
 \mu_i \equiv |E_i|^2 = |\Delta_i|^{-1}.
}
This can be generalized to imaginary images as 
\eq{
 \mu_i = |\Delta_i|^{-1} \expo{2 \nu h_i}
 \label{eq:mu}
}
where $h_i$ is the real part of the phase $\ic \phi_i$. 
From their definitions, one can see that $h_i < 0$ for an effective imaginary image and $h_i = 0$ for a real image.
Thus, the imaginary images are exponentially suppressed at high frequencies.

\section{Stokes phenomena}
Stokes phenomena in lensing are associated with the sudden change of the number of effective imaginary images in the parameter space \citep{jow21}.
At a Stokes line, an imaginary image can suddenly become relevant, introducing a jump discontinuity to the electric field computed with the eikonal approximation.
A more fundamental definition of the Stokes phenomena, as provided by the Picard-Lefschetz theory, is the change in the topology of the Lefschetz thimbles.

Here, we systematically study the Stokes phenomena using two lens models as our examples: a rational lens $\psi(x) = 1/(1+x^2)$, and a lens shape 
\eq{
    \psi(x) = \frac{1}{\sum_{k=0}^{k_{\rm max}} \frac{2^{-k} x^{2k}}{k!}} \,.
}
This lens shape approaches a Gaussian $\expo{-x^2/2}$ at $k_{\rm max} \to \infty$. 
We refer to it as the `simG lens' and use $k_{\rm max}=2$ in this paper.
The resulting lens shapes are shown in Fig.\;\ref{fig:psi}.

\subsection{Locating Stokes lines}
Since Stokes lines are associated with the global change of the Lefschetz thimble topology, it is non-trivial to locate them.
Here we demonstrate a way to locate the Stokes lines in the $\kappa-\beta$ parameter space.

We utilize the fact the number of effective imaginary images changes at a Stokes line. 
Since in general the imaginary phase function $H$ along Lefschetz thimbles associated with different images has different constant values, one condition for a change of image number is that the $H$ values of some images coincide and their Lefschetz thimble pieces merge.
Thus, for a certain lens amplitude $\kappa$, we evaluate the $H$ values of all images for different source locations $\beta$, and the Stokes lines must associate with the coincidence of the $H$ values of different images. 
This is the method for locating the Stokes lines in \citet{jow21}.
We note that, however, this is only a necessary condition.
The same condition also applies to caustics.
As we show in Fig.\;\ref{fig:find_stokes}, in the case of a converging rational lens with $\kappa=-10$, the crossing of $H_0$ and $H_{1,2}$ at $\beta=3.67$ is associated with a Stokes line, 
and convergence of $H_{3}$ and $H_{4}$ at $\beta=5.93$ corresponds to a caustic.
It can also happen that the $H$ values of the images cross when their Lefschetz thimble pieces are well separated from each other and thus no Stokes phenomena occur.
The Lefschetz thimble topology must be checked to confirm the Stokes phenomena (Fig.\;\ref{fig:stokes_demo}).


\subsection{Stokes lines as a feature of lens shapes}
Stokes lines are a unique feature of a lens shape in addition to the caustics.
The three panels of Fig.\;\ref{fig:stokes2} show the Stokes lines in the $\kappa - \beta$ parameter space for three lenses: converging and diverging rational lenses, and diverging simG lens with $k_{\rm max}=2$, respectively.
For the rational lens, there are in total five stationary phase points, and the Stokes line network is relatively simple.
For the $k_{\rm max}=2$ simG lens with nine stationary phase points, the Stokes line network is already very complex.
The original analysis of the Stokes lines \citep{jow21} focused on the weak lensing regime, using a specific example of the converging rational lens at $|\beta| < 1$ where the Stokes lines exist only at $\kappa<0.5$.
In \citet{jow23}, the full Stokes line network which extend to large $\kappa$ values was mapped for the converging rational lens (their Figure\;A1).
In general, Stokes phenomena can occur in both the weak and the strong lensing regimes. 
For some lenses, such as the diverging rational lens, Stokes lines are limited to the weak lensing regime, but for others, such as the converging rational lens and the diverging simG lens, they extend to the strong lensing regime and exist even in the regime where multiple real images exist.
The Stokes phenomena in the strong lensing regime for the converging rational lens are shown in Fig.\;\ref{fig:example} and the associated electric field continuity in Fig.\ref{fig:continuity}.

There exists a clear pattern of how the number of effective imaginary images $N_{\rm im}$ changes in the parameter space for the examples given in Fig.\;\ref{fig:stokes2}: 
 $N_{\rm im}$ is either 0 or 2 at $\beta=0$, and specifically, 0 near $\kappa = 0$, and at $\kappa \to \infty$, $N_{\rm im}=2$ for diverging lenses, and 0 for converging ones.
Crossing a fold caustic or a Stokes line both change $N_{\rm im}$ by one.
At $\beta \to \infty$, $N_{\rm im}$ equals the number of poles of the lens.
These patterns of $N_{\rm im}$ at either $\beta=0$ or $\beta \to \infty$ do not depend on the details of the lens and are likely features of generic lens models. 

\subsection{Electric field across Stokes lines}
We present an example of how the Lefschetz thimble topology and the frequency-dependent electric fields change across a Stokes line and a caustic in Fig.\;\ref{fig:example}.
For this converging rational lens with $\kappa=-10$, there exists a Stokes line at $x=3.67$ (see the top panel of Fig.\;\ref{fig:stokes2}) between the situations demonstrated by the left and middle panels of Fig.\;\ref{fig:example}, and a caustic at $x=5.93$ between the middle and right panels.
The electric field as a function of source location for this whole range $3 < \beta < 7$ is shown in Fig.\;\ref{fig:continuity}.

Across a Stokes line, e.g. from $\beta=3$ to $\beta=4$ in Fig.\;\ref{fig:example}, an imaginary image (purple dot) becomes effective. 
It is associated with a piece of Lefschetz thimble that accounts for a part of the electric field (purple dashed line), and significantly changes the thimble piece (blue solid line in the upper panel) and the associated field (blue dashed line in the lower panel) of one neighboring image.
As expected, the electric field contribution of this imaginary image is exponentially suppressed at high frequencies according to Eq.\;\ref{eq:mu}.
The same applies to the effective imaginary image (yellow dot) that comes into play after the caustic crossing at which two real images (orange and green dots) annihilate. 

It is noteworthy that the total electric field is continuous across both Stokes lines and caustics, as is clearly demonstrated by the continuity of the solid lines in Fig.\;\ref{fig:continuity}.
What changes is its distribution among the images/Lefschetz thimble pieces.
In the eikonal approximation, however, the electric field is discontinuous across Stokes lines, although it does not diverge as it does at caustics (dashed lines in Fig.\;\ref{fig:continuity}).
The amplitude of the electric field jump at the Stokes lines in the eikonal approximation increases at decreasing reduced-frequency $\nu$
as the imaginary images become brighter.

Eikonal approximation with the inclusion of imaginary images (black dashed lines in Fig.\;\ref{fig:example}) in general works better than without them (black dotted lines).
When multiple real images are present (e.g. middle panels of Fig.\;\ref{fig:example}), this improvement is usually moderate. 
This is because the real images dominate the electric fields at high frequencies, and the eikonal approximation has limited accuracy at low frequencies when the imaginary images start to be effective.
When there is only one real image (e.g. right panels of Fig.\;\ref{fig:example}), however, it is crucial to consider the imaginary images to reproduce the oscillatory frequency modulation of the electric field. 
Eikonal approximation with the single real image can fail completely for a wide range of observing frequencies between the high-frequency limit where the imaginary images are negligible and the low-frequency limit where diffraction completely smooths out any lensing signature. 
This range of this intermediate frequency where imaginary images make a significant contribution is broader when the imaginary images are closer to the real axis, i.e. at source locations near caustics.

\subsection{Observable signature of Stokes phenomena}
An imaginary image can be very bright when it lies close to the real axis, where it can be observed as a bright spot in the brightness distribution at some observing frequencies (Shi in prep). 
It can be also observed through interference with other images, which is most prominent when only one other image is present e.g. in the weak lensing regime discussed by \citet{jow21}.

The continuous, finite electric fields at the Stokes lines make them impossible to observe from the first method i.e. brightness distribution at a single frequency. 
The only possibility to observe the Stokes phenomena is through the change of interference behavior in the electric field across the Stokes lines, with the interference observed as frequency modulations.
The change of interference is most distinct at Stokes lines separating a region with one real image and that with one real plus one imaginary image, where interference exists only on one side of the Stokes line.
In the three lens examples in Fig.\;\ref{fig:stokes2}, such Stokes lines exist only for weak lenses with $\kappa<0.5$ at small $\beta$ values.
Fig.\;\ref{fig:sign_of_stokes} shows an example of the onset of interference across a Stokes line at $\beta=0.042$ for the converging rational lens with $\kappa=-0.4$.
After crossing the Stokes line, a frequency oscillation in the electric field shows up as a result of the interference between the real and the imaginary images.

\begin{figure*}
    \centering
    \includegraphics[width=0.98\textwidth]{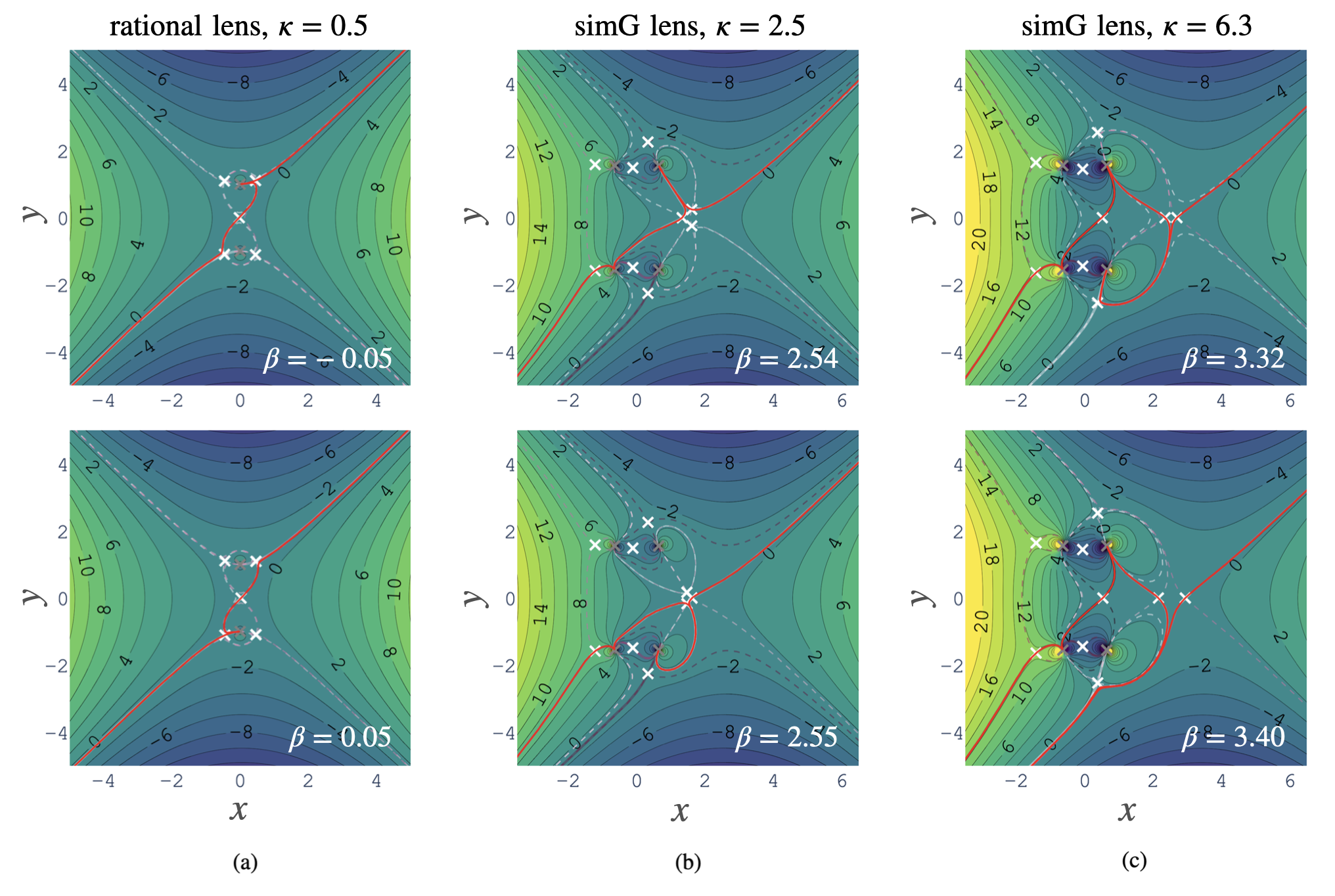}
        \caption{
            \textbf{Examples of high-order Stokes phenomena} where the number of effective images are continuous but the topology of the Lefschetz thimbles changes.
            The three subplots represent regions across three 2nd order Stokes points (a) at $\kappa=0.5$, $\beta=0$ for the diverging rational lens, (b) at $\kappa = 2.58$, $\beta = 2.58$ for the diverging simG lens, and (c) at $\kappa = 6.98$, $\beta = 3.38$ for the diverging simG lens (see Fig.\;\ref{fig:stokes2}).
            The two panels in each subplot show Lefschetz thimble topologies (red line) of two regions in the $\kappa-\beta$ parameter space that are connected at the corresponding 2nd order Stokes point.
            A 2nd order Stokes point is located at the point in the parameter space where 1st order Stokes lines meet, which is also where the $H$ function values (background contours) of more than two images (white crosses) coincide.
            }
        \label{fig:highorder_demo} 
\end{figure*}

\subsection{High-order Stokes phenomena}
When the Stokes phenomena are identified with changes in the topology of Lefschetz thimbles, it is possible to have high-order Stokes phenomena where the system has a continuous number of effective imaginary images but with an abrupt change in the way they are connected to each other by the Lefschetz thimbles.
This could occur when the $H$ values of more than two imaginary images cross at the same location in the parameter space.
In the two-dimensional parameter space of 1D lenses used in this paper, the high-order Stokes phenomena have zero measure and occur only at discrete points where 1st order Stokes lines meet (see Fig.\;\ref{fig:stokes2}), similar to high-order caustics beyond the fold caustic.

Fig.\;\ref{fig:highorder_demo} shows three examples of 2nd order Stokes phenomena. 
The two panels in each subplot show the Lefschetz thimble topologies of two regions in the $\kappa-\beta$ parameter space that are connected at a 2nd order Stokes point.
For the three subplots, the corresponding Stokes points are (a) at $\kappa=0.5$, $\beta=0$ for the diverging rational lens, (b) at $\kappa = 2.58$, $\beta = 2.58$ for the diverging simG lens, and (c) at $\kappa = 6.98$, $\beta = 3.38$ for the diverging simG lens, respectively (see Fig.\;\ref{fig:stokes2}).

A generic crossing of two 1st order Stokes lines divide the two-parameter space into four regions. 
Since regions separated by a 1st order Stokes line must have numbers of effective images differ by one, two of these four regions must have the same number of effective images (i.e. with the same color in Fig.\;\ref{fig:stokes2}).
Crossing from one of them to the other corresponds to a 2nd order Stokes phenomenon.
In these two regions, although the number of effective images are the same, which imaginary images are effective and how they are connected to the real images by the Lefschetz thimbles could change. 
Subplots (a) and (b) of Fig.\;\ref{fig:highorder_demo} show examples of such changes. 
After the crossing of the 2nd order Stokes point, a different imaginary image and a different pole are involved in the Lefschetz thimbles.

Subplot (c) of Fig.\;\ref{fig:highorder_demo} shows a more special case where the 2nd order Stokes point is formed by two 1st order Stokes lines intersecting and terminating at a cusp point which lies on a caustic line.
The two 1st order Stokes lines and the caustic line together divide the parameter space into four regions.
The two regions presented in subplot (c) have not only the same \textsl{number} of effective images but exactly the same effective images. 
The only difference is that, the two real images near $x=2.5$ are connected by a pole in one region, and an infinity in the other.

In a higher-dimensional parameter space e.g. for 2D lens models, Stokes phenomena of higher orders can occur.
Their systematic characterization calls for an analogy of the catastrophe theory for caustics and is beyond the scope of this paper.

\section{Conclusion}
\label{sec:conclusion}
The Stokes phenomena are a fundamental feature of lensing in the wave optics formulation.
The first-order Stokes phenomena are associated with the change in the number of effective imaginary images in the parameter space, 
just as caustic crossings are associated with the change in the number of geometric images.
A more fundamental definition of the Stokes phenomena is the change of the topology of the Lefschetz thimbles, which correspond to discrete lensing images generalized to the wave optics framework. 
High-order Stokes phenomena exist, where the system has continuous number of imaginary images but abrupt changes in the way they are connected to each other by the Lefschetz thimbles.


Using the Picard-Lefschetz theory, we mapped the intricate network of Stokes line and caustics for a few lens models.
Our findings unveiled the occurrence of the Stokes phenomena in both weak and strong lensing regimes.
We also demonstrated how the number of effective imaginary images changes in the parameter space for these examples.

Since the electric field is continuous across the Stokes lines, 
observing the Stokes phenomena is solely achievable through the change of interference behavior in the electric field across the Stokes lines,  
whereas the interference can be observed as frequency modulations.
The onset of frequency oscillation in the electric field is potentially observable in some lensing systems and useful for inferring lens parameters.


\begin{acknowledgments}
\noindent XS thanks the referee for the helpful comments and suggestions. This work is supported by NSFC No. 12373025.
\end{acknowledgments}
\bibliographystyle{aasjournal}
\bibliography{bibliography}


\end{document}